\documentstyle[aps,pra,epsf]{revtex}

\begin{document}

\twocolumn

\newcommand{\bt}{\mbox{\boldmath{$t$}}}
\newcommand{\bs}{\mbox{\boldmath{$s$}}}
\newcommand{\bn}{\mbox{\boldmath{$n$}}}
\newcommand{\bz}{\mbox{\boldmath{$z$}}}
\newcommand{\br}{\mbox{\boldmath{$r$}}}
\newcommand{\bJ}{\mbox{\boldmath{$J$}}}
\newcommand{\hs}{{\hat{s}}}
\title{Cascading Parity-Check Error-Correcting Codes} 
\author{Ido~Kanter$^{1}$ and David~Saad$^{2}$}
 \address{$^{1}$ Department of Physics, Bar-Ilan
University, Ramat-Gan 52900 ,Israel. \\ $^{2}$The Neural Computing Research
Group, Aston University, Birmingham B4 7ET, UK.}  \maketitle
\begin{abstract}
A method for improving the performance of sparse-matrix based parity
check codes is proposed, based on insight gained from methods of
statistical physics. The advantages of the new approach are
demonstrated on an existing encoding/decoding paradigm suggested 
by Sourlas. We also discuss the application of the same method to more
advanced codes of a similar type.
\end{abstract}

Error-correcting codes are commonly used in most means of information
transmission. The coding efficiency, measured in the percentage of
informative transmitted bits, plays a crucial role in determining the
speed of communication channels and the effective storage space on
hard-disks.

The question of channel capacity of noisy communication channels was
addressed by Shannon~\cite{Shannon} in his pioneering work from 1948.
Rigorous bounds have been derived for the maximal transmission rate for
which codes, capable of achieving arbitrarily small error probability,
can be found.

In a typical scenario, a message comprising $N$ binary bits is
transmitted through a noisy communication channel; the received string
differs from the transmitted one due to noise (for instance,
background radiation, defects in magnetic materials, thermal noise
etc) the result of which is the flipping of some bits. Here, we denote
the flipping rate of a bit in a binary symmetric channel (i.e., from 0
to 1 or from 1 to 0) by $f \!\in\! \lbrack 0:1\rbrack$; other types of
noise may also be considered, which may be more realistic in some
scenarios.  Error-correcting codes have been devised for retrieving
the original message at the receiving end.

The receiver can correct the flipped bits (or some fraction of them)
in a retrieved message of length $N$ only if the source transmits
$M(f)\!>\!N$ bits; the ratio between the original number of bits and
those of the transmitted message $R\equiv N/M$ is termed the
code-rate.  Shannon~\cite{Shannon} obtained the optimal trade-off
between the following three quantities in the thermodynamic limit: the
maximal code rate $R_{c}$ (termed channel capacity), the flip rate $f$
(due to noise) and the coded bit error probability $p_b$, given explicitly by
\begin{equation}
\label{eq:shannon_bound}
R_{c}=(1-H_2(f))/(1-H_2(p_b)) , 
\end{equation}
where $H_2(x)=x \ \log_{2}(x)+(1-x) \ \log_{2}(1-x)$.

Shannon's theory indicates the existence of optimal codes but does not
provide a way of obtaining them. Many algorithms were devised to
overcome this practical problem (for a review see\cite{book}),
however, the performance of most algorithms is below Shannon's bound.

One error-correcting code which recently became popular is the
Gallager code~\cite{Gallager,MacKay,Shokrollahi,Richardson}. In this
method, the transmitted message comprises the original message itself
and additional bits used for error correction. Each one of the
additional bits is generated by summing up randomly selected message
bits; the parity of the sum constitutes the transmitted code-word
bit. The choice of the message-vector elements used for generating
single code-word bits is carried out according to a predetermined
random set-up and may be represented by a product of a randomly
generated sparse matrix and the message-vector in a manner explained
below. It was shown that by using Gallager's method it is possible to
get closer to Shannon's bound of the maximal channel capacity for
specific choices of the encoding/decoding
matrices~\cite{MacKay,Shokrollahi,Richardson,us_prl}.

In this paper we will introduce a method for constructing the
encoding/decoding matrix employed in general Gallager-type codes that
enables one to improve the code's performance significantly. In this
method the matrix comprises specifically constructed sparse matrices,
designed to gradually build up the overlap between the original and
the decoded message. We will demonstrate to potential of the method by
examining the model suggested by Sourlas\cite{Sourlas}, representing a
particular case of Gallager's code. The performance of this simple
model is generally inferior to that of other advanced Gallager
codes. However, we prefer to present the main ideas behind our method via
the code of Sourlas\cite{Sourlas} due to its straightforward relation
to Ising spin models and the transparent interpretation of the
construction.  More general constructions have been employed recently
for improving the performance of more complicated Gallager type
codes\cite{us_prl} bringing their performance close to saturating
shannon's bound.

In a general scenario, a message $\bs$ is encoded to a code-word
$\bt$, which is then transmitted through a noisy channel. The
code-word is corrupted during transmission by noise, represented by
the vector $\bn$, and the received code-word $\br$ is decoded by the
receiver for retrieving the original message.

Sourlas's approach is based on mapping the coding problem onto that of
an Ising spin system. The original presentation\cite{Sourlas} made use
of a binary ($\pm1$) message vector representation; here, for brevity
and consistency with notation commonly used for Gallager's method, we
will mostly use the Boolean (0,1) formulation of the problem unless
stated otherwise.  In this approach one constructs code-word bits by
taking the sum of randomly selected $K$ Boolean message bits (mod 2)
\[{\bf t} =  A \ \bs \ \mbox{(mod 2)} \  \]
were the matrix $A$ contains $K (\ll N)$ unit elements per row and $C
(=KM/N)$ per column, setting all other elements to zero.  

Using a simple linear transformation, $\hs_{i}=(2 s_i-1)$, the Boolean
bits $s_i \in (0,1)$ can be mapped onto binary ones $\hs_{i} \in
(-1,1)$. The physical system then consists of $N$ Ising spins with $K$
spin interactions and a fixed connectivity $C$, i.e., each spin
participates in $C$ interactions.  The corresponding Hamiltonian has
the following form:
\begin{equation}
\label{eq:hamiltonian}
H= -\sum_{\langle i_1,i_2 \ldots i_K \rangle}
J_{i_1,i_2,...,i_K} \ S_{i_1}S_{i_2} ...S_{i_K} \ ,
\end{equation}
where $\{S_i \}$ are the binary dynamical variables used in the
decoding process, which can take the values $(\pm 1)$. The noisy free
channel interaction tensor $J^{0}_{i_1,i_2,...,i_K}=
\hs_{i_{1}}\hs_{i_{2}} ...\hs_{i_{K}}$ where $\hs$ is the binary
representation of the originally Boolean message vector $\bs$; the
choice of indices $i_1,i_2,...,i_K$ is predetermined between the
sender and the receiver, reflecting the non-zero row elements of the
matrix $A$. Due to corruption during transmission
$J_{i_1,i_2,...,i_K}= \hs_{i_1}\hs_{i_2} ...\hs_{i_K}$ with
probability $1\!-\!f$ and $-\hs_{i_1}\hs_{i_2} ...\hs_{i_K}$ with
probability $f$.  Under a gauge transformation this model is mapped
onto a highly diluted Ising spin system with ferromagnetic bias (we
assume $f\le 0$).  The magnetization $m=1/N \ \sum_{i=1}^{N} s_{i}
\hs_{i}$ is related to the number of correct bits $(1+m)/2$.

Finding the ground state of the Hamiltonian, in terms of the variables
{\bf S}, corresponds to the Bayes optimal estimation of the original
message bits and thus to decoding the received
message\cite{Sourlas_EPL}; it can be carried out using various
techniques, including energy minimization (simple Monte-Carlo at some
temperature, say zero) or belief propagation
(e.g.~\cite{MacKay,Frey}). The properties of Sourlas's method have
been investigated for the fully connected\cite{Sourlas} and diluted
cases\cite{ks_sourlas} with fixed connectivity.


The main drawback of the method is the need to compromise between
superior capabilities and poor decoding performance, and sub-optimal
capabilities (in terms of the achievable code rate) and successful
decoding. For example, the choice of $K\!=\!2$ in Sourlas's approach,
i.e., having only 2 Multi-Spin Interactions (MSI), corresponds to an
energy landscape dominated by a very large basin of attraction; this
will lead to a successful convergence from almost {\em any} small
positive initial overlap between the dynamical variables and the
message, and will result in a large end-overlap $m$ (and consequently
successful decoding). However, this overlap is much smaller than 1,
the {\em perfect} decoding required in most cases, even for flip rates
way below Shannon's bound.  On the other hand, choosing higher $K$
(and consequentially higher $C$) values may result in very high
end-magnetization and successful decoding, but also in a corresponding
dramatic decease in the basin of attraction. The improved
end-magnetization can be easily understood as the increased
connectivity reduces the probability of a negative local field
asymptotically; the reduced basin of attraction clearly results from
the vanishing contribution of the product of $K$ spins far from the
ground state.  Consequently, one may expect a decoding failure unless
the starting point is chosen very close to the original message; such
information is clearly unavailable in practical scenarios. One should
emphasize that the basin of attraction shrinks dramatically, for
instance, for $K\!=\!6$ the initial magnetization required for
successful convergence is higher than $0.98-0.99$.

Our method builds on insight gained from the study of physical systems
with symmetric and asymmetric\cite{ido} multi-spin interactions and
previous studies of Sourlas's code via methods of statistical
physics\cite{ks_sourlas}.  It is based on the gradual introduction of
higher connectivity sparse matrices, exploiting the excellent
convergence properties of codes based on low $K$ values with the high
performance of high-$K$ codes. For example, one may carry out the
first stage of the decoding process using $K\!=\!2$ and then, once the
overlap between the decoded word and the original message is within
the relevant basin of attraction, one invokes the $K\!=\!3$
connections (that were already used in generating the code-word),
resulting in a much higher overlap in comparison to the case in which
only $K\!=\!2$ connections are used. The process can clearly be
generalized to include a longer sequence of transitions and to
different $K$-values, such as to improve the overall performance. One
should point out that from a physical point of view this is equivalent
to changing the Hamiltonian (\ref{eq:hamiltonian}) as a function of
time.

It has been shown that the method does not have to be implemented in a
dynamical manner as the one described above as long as the
encoding/decoding structured matrix is constructed appropriately. The
dynamical implementation is slightly superior close to the critical
flip rate. It also enables one to obtain, at zero temperatue, results
which are typically obtained only at finite temperatures.

An optimal construction of the encoding/decoding matrix is clearly the 
key point to a successful algorithm. Although there is no clear recipe 
for constructing the matrix in general, one can provide a few guidelines
that are helpful for improving the performance.

Originally, the method relies on invoking the next level connections
once the current state of the system (and the resulting overlaps with
the input message) is within its basin of attraction. The latter can
be estimated numerically either by an exhaustive search or
approximated analytically, by considering contributions to a single
node and averaging over the input probabilities.  This approximation
assumes magnetization $m$ per contributing node, where $\langle S_i(t)
\rangle =m$ for all $i$ neglecting correlations among the different
sites (spins). The average $\langle \cdots \rangle$ represents an
average over possible spin distributions and weights, where the prior
on the weights is taken as $P(J) = f \ \delta(J\!+\!1) +(1\!-\!f) \
\delta(J\!-\!1)$ and $P(m) = (1\!+\!m)/2 \ \delta(m\!-\!1)
+(1\!-\!m)/2 \ \delta(m\!+\!1)$.  The basin of attraction at zero
temperature is then defined as the minimal magnetization such that
$\langle \mbox{sign}(h)\rangle \ge m$, where $h$ is the induced
field. The end magnetization obtained is defined as the $m$ value for
which the equality holds. To demonstrate the agreement between results
obtained numerically and analytically and how they reflect the essence
of the cascaded decoding approach, we examine analytically the case of
$R\!=\!1/3$ and $f\!=\!0.14$: (a) Having only 2 MSI and $C\!=\!6$
($=\!K/R$) interactions one obtains a value of $m\!=\!0.932$ for the
end magnetization in comparison to $m \approx 0.94$ obtained
empirically; no significant limitation on the basin of attraction has
been observed, i.e, convergence is expected from any initial overlap.
(b) For ($C\!=$) 4 interactions with ($K\!=$) 2 MSI and $C\!=\!4$
interactions with 4 MSI (this provides in total the same message
length as before $1/R\! =\! 3\! =\! 4/2 + 4/4$) one obtains an end
magnetization of $m\!=\!0.97$; the value obtained empirically is $m
\approx 0.98$. The basin of attraction requires an initial overlap $m
\ge 0.6-0.62$. (c) For only 4 interactions with 2 MSI (i.e., as in (b)
but omitting all 4 MSI components) one obtains a final magnetization
of $m\!=\!0.64$ with no restrictions on the basin of attraction.

This forms the basis for the cascading error-correcting method:
Starting from any (positive) initial overlap between the message and
the dynamical variables, and employing the configuration of (c), one
obtains an end magnetization which is well within the basin of
attraction of the complete Hamiltonian system (b). Following with the
dynamics of the complete system results in an end magnetization of $m
\approx 0.98$, well above the end magnetization of (a), although the two
systems have the same code-rate and initial basin of attraction.

The optimal combination of MSI, for which the highest end
magnetization is obtained, depends on many parameters including the
code rate $R$, the noise level and the message bias.  Finding the
optimal connectivities ratio of two different MSI values can be
carried out by plotting the end magnetization of a partial system with
only low MSI connections (such as (c) with $C\!=\!4$ and $K\!=\!2$)
against the minimal magnetization required for convergence (basin of
attraction) in the complete system (as in
(b)). Figure~\ref{fig:mag_frac} shows the two curves as a function of
the fraction $0 \le \rho \le 1$ of $K\!=\!4$ interactions (and with
$(1-\rho)$ $K\!=\!2$ interactions). The experiment has been carried
out for the case of $R\!=\!1/3$, $f\!=\!0.14$ and $N\!=\!10^{4}$ and
the results were averaged over 10 trials. From
figure~\ref{fig:mag_frac} it is clear that any choice of $\rho \le
0.42$ will lead the partial configuration to a higher end
magnetization than required for the complete system to converge. As
systems with higher connectivity will result in higher
end-magnetization, one should aim at choosing the highest $\rho$ value
for which the partial system's end-magnetization is higher than the
basin of attraction of the complete system, i.e., the intersection of
the two graphs.

In constructing the matrix one may have to use a non-integer effective
number of connections per spin $C$ (i.e., the number of non-zero
elements in a column of the matrix $A$) due to its relations to other
system parameters; denoting the number of message bits with $K_k$ MSI as
$N_k$ than $\sum_k N_k = 1/R$ and $C = \sum_k K_k N_k/N$ where $N$ is
the number of message elements.  We have found it useful to keep the
distribution of non-zero elements per column as homogeneous as
possible to provide equal corrective contribution to all bits.
In addition, it would be helpful to avoid having small loops in the
connectivity matrix, i.e., small groups of sites connected cyclically,
as these contribute to recurrent dynamics which suppresses corrective
input from the rest of the system.

For converging to the correct final state it is useful to initialize
the system with some positive overlap between the dynamical variables
and the original message as in most of these systems, both solutions,
with $m=\pm1$, are equally attractive. This may be achieved either by
transmitting a small fraction of the message itself simultaneously,
which is the less favorable solution, or by adding some odd-MSI (i.e.,
an odd $K$ value, e.g., $K\!=\!3$) to the mainly even $K$ value used
initially (e.g., $K\!=\!2$); this assists in breaking the symmetry
from any initial dynamical variables setting with practically no
effect on the basin of attraction.  Odd connectivity per spin also
alleviates the problem of zero local field; it is characterized by
finite zero temperature entropy and hence improves convergence and the
performance in general.

As decoding is carried out iteratively, it is important to define
a halting criterion for obtaining the decoded message.  Here we carry
out a simple energy minimization, and the algorithm comes to a halt
when all spins are aligned to their local fields (except spins of zero
local fields). The local field of each spin $h_{i}$ is
calculated in turn by summing over all other spin states
\[ h_{i} = \sum_{\langle i_2 \ldots i_K \rangle}
J_{i,i_2,...,i_K} \ S_{i_2} ...S_{i_K} \  . \]
The binary value of the individual spin is then obtained by aligning
it with the value calculated for $h_{i}$, i.e., $S_{i}=+1$ if
$h_{i}>0$ and $S_{i}=-1$ otherwise. This dynamical process is repeated
until the system stabilizes or until some halting criterion is obeyed.

To show the excellent performance of the new method we compared the
end-overlap of four different systems of rate $R\!=\!1/7$ (i.e.,
code-word length of $M\!=\!7N$, corresponding to a critical flip rate
$f_c\!=\!0.282$ due to Shannon, see Eq.(\ref{eq:shannon_bound})),
starting from similar initial conditions and the same transmission
flip rate $f\!=\!0.25$. Experiments have been carried out for
different system sizes, $N\!=\!5000-20000$, for getting a feel for 
the dependence of the performance on the system's size.  The dynamics
employed for the energy minimization is based on sequentially updating
the spins although similar results were obtained for parallel
dynamics.  The results summarized in table~\ref{tab:sourlas} show a
significant improvement in the final overlap due to the cascaded
encoding/decoding scheme. The low MSI part of the code word serves to
bring the decoded vector to a sufficiently high overlap with the
original message, so that it lies within the basin of attraction of the
combined message; the final convergence to a very high overlap is
facilitated by the high MSI.

These results, can be improved upon by having a reliable prior
knowledge of the noise level.  No such knowledge was assumed in any of
the above mentioned experiments.  One should also point out that in
all the experiments, we observed convergence after a few tens of
iterations at most and the complexity of the algorithm used is of
$O(N)$. The physical interpretation to the success of
this method benefits from viewing the system as a graph whereby the
different nodes (message sites) are sparsely connected by unit weights
(elements of the multidimensional tensor $\bJ$). By increasing the
number of MSI, say from $K\!=\!2$ to $K\!=\!4$, one increases the
graph connectivity and the number of inputs which contribute to
determining the state of each specific spin. On the other hand, as the
number of MSI increases, their `quality' deteriorates; the average
local field is determined by taking the product of $K\!-\!1$ terms
representing the magnetization of all spins connected to the relevant
weight, which decays rapidly with the increase in MSI (e.g., a
magnetization of 0.6 in the case of $K\!=\!2$ will reduce to
$0.6^{3}\!=\!0.216$ in the case of $K\!=\!4$).  Furthermore,
increasing the number of MSI creates exponentially many local minima
and energy levels which are highly degenerate (for instance, one local
field obtained in a system with $K$ MSI has a degeneracy of
$2^{K-1}$), making the probability of successful convergence
vanishingly small.  This interplay is at the center of our approach
and guides the choice of the optimal model parameters.

To conclude, we have shown that through a successive change in MSI and
connectivity one can boost the performance of matrix based
error-correcting codes. We showed that this is feasible for the
special case of Gallager-type codes presented by
Sourlas\cite{Sourlas}, although the method itself is applicable to all
codes of this type and may be easily adapted to fit most of the
existing variations as will be shown elsewhere~\cite{us_prl}.  There
are a few extensions that we would like to point out: \\
1) Although our examples concentrated on unbiased messages, the
process can clearly be easily generalized to biased messages. It may
also be generalized to include non-symmetric connections and
continuous or multilevel message units (instead of binary).
\\
2) It is plausible that many sets of parameters have similar
performance in the thermodynamic limit, however, their finite size
behavior above and below saturation is of great interest from a
practical point of view. Finding architectures that are superior in
their finite size behavior, as well as finding methods to suppress the
finite size effects, would clearly be of great practical significance.

The cascading decoding method and the extensions mentioned above open
a wide range of possibilities for a highly efficient encoding/decoding
mechanism of significant practical value.

\vspace*{-2mm}
\begin{table}
\begin{center}
\begin{tabular}{|c|c|c|} 
Message bits ($N_k$) &  MSI ($K_k$)   &  Final magnetization \\ \hline \hline 
$7N$  &  $2$  &  0.940  \\ \hline 
$5N$  &  $2$  &  0.975  \\ 
$2N$  &  $5$  &    \\ \hline 
$5N$  &  $2$  &  0.987  \\ 
$2N$  &  $7$  &    \\ \hline 
$5N$  &  $2$  &  0.993  \\ 
$2N$  &  $9$  &   \\   
\end{tabular}
\vspace*{1mm}
\caption{Final overlap for various combinations of MSI in the case of
$R=1/7$ and transmission flip rate of $f=0.25$, starting from similar
initial conditions for the different configurations.}
\label{tab:sourlas}
\end{center}
\end{table}

\vspace*{-1.2cm}

\begin{figure}
\begin{center}
\epsfysize = 6.0cm
\epsfbox[0 120 670 520]{Sour_opt.eps}
\end{center}
\vspace*{0.5cm}
\caption{The end magnetization of a partial system (diamonds) with
only low MSI connections ($C\!=\!4$ and $K\!=\!2$ and $M=3N$) against
the minimal magnetization required for convergence (basin of
attraction) in the complete system (triangles). The two curves are
plotted as a function of the fraction $0 \le \rho \le 1$ of $K\!=\!4$
connection (and $(1-\rho)$ connections of $K\!=\!2$).}
\label{fig:mag_frac}
\end{figure}

\end{document}